\documentclass[10pt,pre,twocolumn]{revtex4-1}

\usepackage{dcolumn}						
\usepackage{bm}							

\usepackage{graphicx} 
\usepackage{amssymb}
\usepackage{latexsym}
\usepackage{hyperref}
\usepackage{color}

\newcommand \be {\begin{equation}}
\newcommand \ee {\end{equation}}
\newcommand \bea {\begin{eqnarray}}
\newcommand \eea {\end{eqnarray}}
\newcommand \mC {\mathcal{C}}
\newcommand \ve {\varepsilon}

\newcommand{\la}{\langle}
\newcommand{\ra}{\rangle}
\newcommand{\fh}{f_{\mathrm{h}}}
\newcommand{\fs}{f_{\mathrm{s}}}

\begin{document}

\title{Statistical physics of frictional grains: some simple applications of Edwards statistics}

\author{Eric Bertin}
\affiliation{LIPHY, Universit\'e Grenoble Alpes and CNRS, 38000 Grenoble, France}

\begin{abstract}
Granular matter like sand is composed of a large number of interacting grains, and is thus expected to be amenable to a statistical physics treatment. Yet, the frictional properties of grains make the statistical physics of granular matter significantly different from the equilibrium statistical physics of atomic or molecular systems.
We illustrate here on simple models some of the key concepts of the statistical physics introduced by Edwards and coworkers more than thirty years ago to describe shaken granular piles.
Quite surprisingly, properties of such frictional systems observed at high effective temperature (i.e., strong shaking) may share some analogies with some low temperature properties of equilibrium systems.
For instance, the effective specific heat of non-interacting frictional grains under strong shaking in a harmonic potential goes to zero in the high temperature limit.
As a second example, a chain of frictional grains linked by springs exhibits a critical point at infinite effective temperature, at odds with the zero-temperature critical point generically found in one-dimensional equilibrium systems in the presence of local interactions.
\end{abstract}

\maketitle

\section{Introduction}

Whoever has played with sand on a beach may have been amazed by the mechanical properties of sand. If you take some dry sand in your hands and gently open fingers, sand flows almost like water. Yet, the beach forms an essentially solid surface on which you can walk, leaving some footprints that are only partially erased by the spontaneous relaxation dynamics of dry sand.
One may thus wonder how materials like sand that are formed of grains of macroscopic size (and thus called `granular materials') can exhibit mechanical properties that are intermediate between solids and liquids.
The reason for these non-standard properties of granular materials lies in the frictional properties of grains. Two grains in contact experience dry friction, which means that they can support a certain amount of tangential forces at contact without gliding. This property is no longer true with viscous friction. Simply think of beads covered with oil and put in contact: any tiny amount of tangential forces would make them glide one on top of the other.

On the other side, the fact that granular materials are made of a large number of grains naturally calls for a statistical description.
However, this statistical description is expected to differ from the equilibrium statistical physics formalism that describes materials made of atoms or molecules.
While atoms or molecules have a conservative dynamics, the frictional properties of grains make their dynamics dissipative, which deeply modifies their large-scale statistical properties. In more abstract terms, the time reversal symmetry is broken in granular materials.
This important difference of granular materials with respect to equilibrium systems has been taken into account through a minimal generalization of the equilibrium statistical physics framework, as proposed by Edwards and coworkers in the late 1980's \cite{Edwards,ME89}.
The goal of this paper is to illustrate the Edwards theory of granular matter on simple and pedagogical examples.

The Edwards theory notably introduces an effective temperature as a parameter characterizing the statistics of the packing of grains.
It is also related to the amount of injected power in the packing through external forces, for instance by shaking the grains.
This effective temperature is many orders of magnitude larger than the thermodynamic temperature of the material the grains are made of.
Thermal fluctuations associated with the thermodynamic temperature are thus negligible, as they are for instance completely unable to lift a grain over a height of the order of its diameter.

The article is organized as follows. Sec.~\ref{sec:Edw} briefly introduces the general framework of Edwards statistical mechanics for systems with dry friction.
Sec.~\ref{sec:frict:HO} introduces a simple model of non-interacting frictional particles attached to a spring, which generalizes the harmonic oscillator model of statistical mechanics.
The interest of this model is mostly pedagogical, and qualitative analogies with some aspects of quantum harmonic oscillators are outlined.
Then Sec.~\ref{sec:spring:block} discusses a more complicated model of frictional particles linked by springs. The presence of interactions between particles generates strong correlations in a high temperature regime, at variance with usual equilibrium systems where correlations appear at low temperature.
Finally, Sec.~\ref{sec:projects} suggests possible computational projects for students, of varying difficulties, consisting in simulating one of the above models.
Sec.~\ref{sec:conclusion} eventually concludes the paper.

\section{Edwards approach for shaken granular matter}
\label{sec:Edw}

A statistical description of a granular pile is meaningful if the pile is able to visit many different configurations.
To do so, it is convenient to inject energy through a `tapping' protocol, by which the pile is repeatedly shaken and then let relax (after switching off the shaking mechanism) to a mechanically stable configuration, also called blocked configuration for short. A blocked configuration is such that the sum of all forces acting on any given grain is equal to zero.
Thanks to the tapping protocol, many different blocked configurations are visited, and the pile can be described by a statistics of blocked configurations. This statistics should allow for instance for the prediction of average values of observables like the height of the pile or the force exerted by the grains on the container.

Let us call $\mC$ the configuration of the pile, that is the list of all grain positions.
The Edwards approach first postulates that all configurations that are not mechanically stable have zero probability. This is justified by the tapping protocol, in which one registers the successive blocked configurations that are selected by the dynamics after shaking and relaxation.
Then, taking inspiration from equilibrium statistical mechanics, the idea of Edwards and coworkers \cite{Edwards,ME89} is to assume that the statistics of blocked configurations takes the simplest possible form, given the macroscopic constraints to be taken into account.
By analogy with the equilibrium canonical ensemble, one may assume that the granular pile should be described by the most likely probability distribution of blocked configurations with a given average value of the total energy. Here energy may correspond to the potential energy associated with gravity and possibly to elastic contributions.
This assumption leads to the following form of the probability distribution of blocked configurations
\cite{Edwards,ME89,BKVS00} (see \cite{BHDC15} for a review),
\be \label{eq:Edwards:dist:E}
P(\mC) = \frac{1}{Z}\, \exp\left(-\frac{E(\mC)}{T_{\rm eff}} \right) \, \mathcal{F}(\mC)\,,
\ee
where $Z$ is a normalization factor determined by the condition $\sum_{\mC} P(\mC)=1$.
The parameter $T_{\rm eff}$ plays a role similar to the thermodynamic temperature $T$ in equilibrium systems, or more precisely to $k_{\rm B}T$, where $k_{\rm B}$  is the Boltzmann constant. For these reasons, $T_{\rm eff}$ is called an effective temperature, although it has the dimension of an energy.
The indicator function $\mathcal{F}(\mC)$ has been included to select blocked configurations among all possible configurations, that is among all possible positions of the grains.
For a blocked configuration $\mC$, $\mathcal{F}(\mC)=1$, whereas $\mathcal{F}(\mC)=0$ if $\mC$ is not mechanically stable. Although this general definition is simple, the explicit form of the function $\mathcal{F}(\mC)$ may be quite complicated in practice \cite{BE03,BE06,BSWM08,WSJM11,APF14}.
Note that the presence of the indicator function $\mathcal{F}(\mC)$ in the probability distribution of configurations is precisely what makes it different from the equilibrium canonical distribution. In other words, the function $\mathcal{F}(\mC)$ is expected to be the key ingredient allowing for the description of the peculiarities of the granular phenomenology.

Other global observables than the energy may also be taken into account. For instance, looking at the most likely probability distribution of blocked configurations with given average values of the total energy and of the total volume, one finds
\be \label{eq:Edwards:dist:EV}
P(\mC) = \frac{1}{Z}\, \exp\left(-\frac{E(\mC)}{T_{\rm eff}}-\frac{V(\mC)}{X}\right) \, \mathcal{F}(\mC)\,.
\ee
The parameter $X$ is called compactivity, and $X^{-1}$ is the analogue of the ratio $p/k_{\rm B}T$ at equilibrium, where $p$ is the pressure.
Generalizations have also been proposed, taking into account other observables like mechanical stress \cite{HHC07,HC09,BE09,BJE12,BZBC13}.
Here we focus on the simplest case and consider throughout the paper the distribution given in Eq.~(\ref{eq:Edwards:dist:E}) that involves only the energy.

Both experimental \cite{NKBJN98,SGS05,LCDB06,NRRCD09} and numerical \cite{KM02,M04,MD05,PCN06,BK15} tests of the Edwards probability distribution have been performed, using different forms of the probability distribution $P(\mC)$ like Eqs.~(\ref{eq:Edwards:dist:E}) and (\ref{eq:Edwards:dist:EV}). These tests confirm that the Edwards distribution captures at least qualitatively most of the phenomenology of granular matter.
However, a quantitative assessment of the predictions of Edwards theory is made difficult by the complexity of the function
$\mathcal{F}(\mC)$ describing blocked configurations \cite{BE03,BE06,BSWM08,WSJM11,APF14}, which can most often be evaluated only through rather strong approximations \cite{SL03,BE03}.
It is then hard to disentangle the discrepancies resulting from the Edwards prescription and that resulting from the approximations made in the evaluation of the function $\mathcal{F}(\mC)$, when comparing predictions with empirical results.

\section{Non-interacting frictional harmonic oscillators}
\label{sec:frict:HO}

\subsection{Model and dynamics}

Equilibrium statistical physics lectures usually start by describing the simplest possible examples of systems amenable to a statistical treatment, namely systems consisting of a large number of non-interacting particles
\cite{McQuarrie,Hill,Chandler}.
These include the ideal gas, as well as assemblies of non-interacting harmonic oscillators that may be interpreted for instance as the Einstein model of a crystalline solid \cite{Ashcroft}.

\begin{figure}[t!]
\centering\includegraphics[width=6cm]{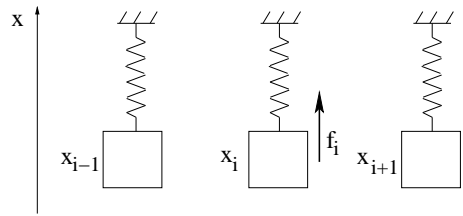}
\caption{Sketch of the frictional harmonic oscillators model, emphasizing the driving force $f_i$.}
\label{fig-HO}
\end{figure}

To illustrate the statistical physics of frictional particles proposed by Edwards and coworkers, we now describe a model of $N$ non-interacting harmonic oscillators with dry friction (Fig.~\ref{fig-HO}).
A physical realization of such a frictional harmonic oscillator is a mass $m$ moving on a horizontal substrate (e.g., a table) with dry friction coefficient $\mu$, and attached to a spring of stiffness $k$, the other end of the spring being attached to a wall. Assuming for the sake of simplicity that the mass $i$ ($i=1,\dots,N$) is constrained to move along a given axis $x$, 
the position $x_i$ of mass $i$ evolves according to
\be \label{eq:dyn:HOfrict}
m \frac{d^2 x_i}{dt^2} = -k x_i - \mu mg\, {\rm sign} \left( \frac{d x_i}{dt} \right) + f_i(t)\,.
\ee
The first term on the rhs of Eq.~(\ref{eq:dyn:HOfrict}) corresponds to the force $-k x_i$ exerted by the spring, assuming that $x_i=0$ corresponds to the rest position of the spring.
The second term corresponds to the dynamic dry friction force, equal to minus the friction coefficient $\mu$ times the weight $mg$ (with $g$ the gravity intensity), times the sign of the velocity. The absolute value of the force is thus independent of the speed, at variance with viscous friction.
Finally, the last term $f_i(t)$ is a driving force used to inject energy during the tapping protocol. It could model, for instance, a horizontal vibration of the table on which the masses are placed.

When at rest ($d x_i/dt=0$), the mass is subjected to a static dry friction force instead of the dynamic dry friction force
appearing in Eq.~(\ref{eq:dyn:HOfrict}).
The static friction force exactly compensates the other horizontal forces as long as these forces do not overcome in absolute value a threshold force equal to $\mu mg$.  
(Note that in realistic systems, the static friction coefficient defining the threshold force slightly differs from the dynamical one appearing in the dynamic friction force, but we neglect this slight difference here for the sake of simplicity.)
When the threshold is exceeded, the mass starts to move.
In the absence of driving force ($f_i=0$), the only horizontal force apart from the static friction force is the force exerted by the spring.
Hence the mass starts to move if $k |x_i| > \mu mg$.
Conversely, blocked configurations correspond to static configurations that verify the condition $k |x_i| < \mu mg$. In other words, blocked configurations satisfy $|x_i| < a$, with a characteristic length scale
\be \label{eq:def:a}
a = \frac{\mu mg}{k}\,.
\ee
Note that the length scale $a$ is proportional to the friction coefficient $\mu$, so that $a$ is nonzero for frictional systems only.

\subsection{Statistics of blocked configurations and average energy}

We now turn to the determination of the stationary probability distribution $P_N(x_1,\dots,x_N)$ associated with the set $(x_1,\dots,x_N)$ of positions of the $N$ masses, using the Edwards prescription given in Eq.~(\ref{eq:Edwards:dist:E}). Here the microscopic configuration $\mC$ appearing in Eq.~(\ref{eq:Edwards:dist:E}) is the list of all particle positions, $\mC=(x_1,\dots,x_N)$. The total energy of a configuration is given by
\be \label{eq:E:HO}
E(x_1,\dots,x_N) = \sum_{i=1}^N \frac{1}{2} k x_i^2\,.
\ee
The indicator function $\mathcal{F}(\mC)$ appearing in Eq.~(\ref{eq:Edwards:dist:E}) can be formally written as
\be \label{eq:F:HO}
\mathcal{F}(x_1,\dots,x_N) = \prod_{i=1}^N \Theta(a-|x_i|)
\ee
where $\Theta(x)$ is the Heaviside function, $\Theta(x)=1$ for $x\ge 0$ and $\Theta(x)=0$ for $x< 0$.
The expression of $\mathcal{F}(x_1,\dots,x_N)$ given in Eq.~(\ref{eq:F:HO}) simply means that a configuration
$(x_1,\dots,x_N)$ is a blocked configuration if all $x_i$ satisfy $|x_i| < a$.
It follows from Eqs.~(\ref{eq:E:HO}) and (\ref{eq:F:HO}) that the Edwards distribution Eq.~(\ref{eq:Edwards:dist:E}) factorizes as
\be
P_N(x_1,\dots,x_N) = \prod_{i=1}^N p(x_i)
\ee
with a one-body distribution $p(x)$ given by
\be
p(x) = \frac{1}{Z_1} \, e^{-\beta k x^2/2} \qquad {\rm if} \quad |x|<a\,, \\
\ee
and $p(x) = 0$ otherwise, with the notation $\beta =1/T_{\rm eff}$.
The quantity $Z_1$ defined as
\be
Z_1 = \int_{-a}^a dx \, e^{-\beta k x^2/2}\,,
\ee
plays the role of a one-body partition function.
The average energy of the $N$ harmonic oscillator system is then obtained as
\be
\langle E \rangle = \frac{N}{Z_1} \int_{-a}^a dx \, \frac{k}{2} x^2 \, e^{-\beta k x^2/2}\,.
\ee
It is convenient at this stage to introduce the characteristic effective temperature 
\be \label{eq:def:Tstar}
T^*=\frac{1}{3}k a^2
\ee
(the reason for introducing the $\frac{1}{3}$ factor is discussed below).
Note that the temperature scale $T^*$ originates from the presence of static friction, as can be seen from the expression of $a$ given in Eq.~(\ref{eq:def:a}).

\begin{figure}[t!]
\centering\includegraphics[width=8cm]{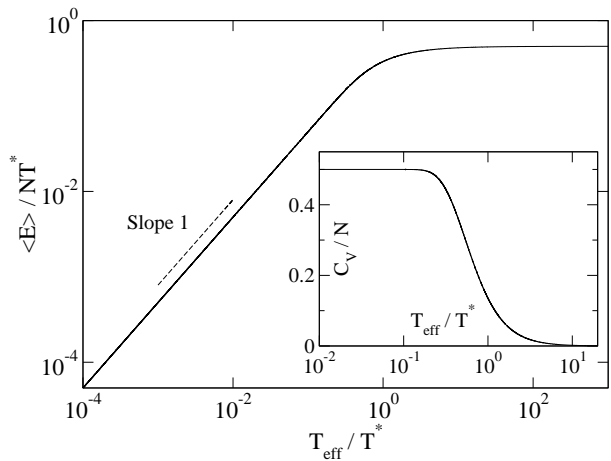}
\caption{Dimensionless energy density $\la E\ra/NT^*$ in the frictional harmonic oscillator model, as a function of the reduced temperature $T_{\rm eff}/T^*$, showing the energy saturation at high temperature. Inset: specific heat $C_V/N$ versus $T_{\rm eff}/T^*$, showing a drop from the classical value $1/2$ (equivalent to the Dulong and Petit law) at low temperature to zero at high temperature.}
\label{fig-energy-HO}
\end{figure}

Thanks to the change of variable $y=\sqrt{\beta k} \, x$,
the average energy $\langle E \rangle$ can be rewritten in the form
(see Fig.~\ref{fig-energy-HO}):
\be
\langle E \rangle = \frac{1}{2} \, N T_{\rm eff} \,
\fh \! \left( \frac{T_{\rm eff}}{T^*} \right),
\ee
having defined the auxiliary function $\fh(u)$ as
\be \label{eq:def:fu}
\fh(u) = \frac{\int_0^{\sqrt{3/u}} dy \, y^2 \, e^{-y^2/2}}{\int_0^{\sqrt{3/u}} dy \, e^{-y^2/2}} \,.
\ee
From this integral expression of the function $\fh(u)$, its asymptotic behaviors can be determined.
One finds $\fh(u) \to 1$ for $u \to 0$ as well as $\fh(u) \approx 1/u$ when $u\to \infty$.

The average energy $\langle E \rangle$ then satisfies generalized equipartition relations
in the two asymptotic regimes $T_{\rm eff} \ll T^*$ and $T_{\rm eff} \gg T^*$:
\bea
\label{eq:equip1:HO}
\langle E \rangle &\approx& \frac{1}{2} \, N T_{\rm eff}  \qquad {\rm for} \quad T_{\rm eff} \ll T^*,\\
\label{eq:equip2:HO}
\langle E \rangle &\approx& \frac{1}{2} \, N T^*   \qquad\; {\rm for} \quad T_{\rm eff} \gg T^*.
\eea
It follows that for a low effective temperature, a form analogous to the equilibrium equipartition relation is obtained, in the sense that the average energy is proportional to the effective temperature (we recall that at equilibrium, the average energy of $N$ harmonic oscillators satisfies $\langle E \rangle = \frac{1}{2} \, N k_{\rm B} T$).
In contrast, for a high effective temperature, the average energy reaches a maximum value, settled by the dry friction coefficient
---see Eqs.~(\ref{eq:def:Tstar}) and (\ref{eq:def:a}).
Note that the $\frac{1}{3}$ factor in the definition (\ref{eq:def:Tstar}) of $T^*$ has been introduced to get an equipartition-like form of Eq.~(\ref{eq:equip2:HO}).

While the current problem is fully within the realm of classical physics, it is of interest to note some analogy with quantum mechanics in the following sense.
Quantum harmonic oscillators have discrete energy levels, and this discreteness leads at low temperature to a concentration of energy on the lowest energy level, with an exponentially small amount of energy on excited levels. As a result, the specific heat, instead of being constant as in classical harmonic oscillators, strongly decreases when decreasing temperature in the low temperature regime. It even goes to zero in the zero temperature limit \cite{Ashcroft}.
At a qualitative level, a somewhat similar phenomenon occurs in the present frictional harmonic oscillator model, when now considering the limit of high effective temperature. Instead of having discrete energy levels, the frictional harmonic oscillator has a bounded continuum of energy levels, where the energy upper bound equal to $T^*/2$ per oscillator originates from static friction.
At low effective temperature ($T_{\rm eff} \ll T^*$), the effect of the energy bound is negligible, and one recovers a constant specific heat
$C_V = d\la E\ra/dT_{\rm eff} = \frac{1}{2}N$ as in classical equilibrium harmonic oscillators, corresponding in the latter case to the Dulong and Petit law \cite{Ashcroft} (setting the Boltzmann constant $k_B=1$). Conversely, in the high temperature regime, the average energy saturates to the upper energy bound, and the specific heat decreases and eventually goes to zero. Using the small-$u$ expansion $\fh(u)\approx u-\frac{2}{5}u^2$
of the function $\fh(u)$ defined in Eq.~(\ref{eq:def:fu}), we obtain
$C_v \approx \frac{1}{5}(T^*/T_{\rm eff})^2$ for $T_{\rm eff} \gg T^*$
(see inset of Fig.~\ref{fig-energy-HO}).

\section{A shaken spring-block model}
\label{sec:spring:block}

\subsection{Model and dynamics}

\begin{figure}[t!]
\centering\includegraphics[width=8cm]{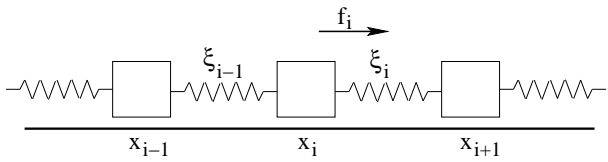}
\caption{Sketch of the frictional spring-block model.}
\label{fig-spring}
\end{figure}

We now go beyond the above non-interacting case, and turn to a second example of a frictional model, to explore further the interesting phenomenology emerging from dry friction.
We consider a one-dimensional chain of $N+1$ frictional blocks of mass $m$ on a substrate. Each block $i=0,\dots,N$, located at position $x_i(t)$, experiences dry friction from the substrate. Blocks are connected by springs of stiffness $k$ and rest length $l_0$, as illustrated on Fig.~\ref{fig-spring}. Springs thus induce interactions between blocks.
This model has been initially introduced in the context of earthquake
modeling \cite{BK67,CL89}, and further studied later on in Refs.~\cite{BPG11,Gradenigo15}.
To avoid taking care of the no-crossing condition between blocks, we assume the rest length $l_0$ to be significantly larger than the typical value of spring extensions.
Similarly to the frictional harmonic oscillator model, a block at rest starts moving if the total force exerted on it by neighboring springs overcomes a threshold force equal to the weight $mg$ times the static friction coefficient $\mu$. If the force exerted by springs remains below the threshold force, the block does not move.
We can thus define a blocked configuration as a configuration of the positions $x_i$, $i=0,\dots,N$, such that resulting spring forces on each block do not exceed the threshold value $\mu mg$.

Blocked configurations are sampled using a driving protocol that periodically injects energy into the system.
Each period is decomposed into a driving phase of duration $\tau$, during which a strong external force is applied to each block.
The driving phase is followed by a relaxation phase during which the system relaxes to a blocked configuration.
The driving and relaxation phases are described by the following dynamics.
\be \label{eq:dyn:spring-block}
m \frac{d^2 x_i}{dt^2} = - \mu mg\, {\rm sign}\left( \frac{d x_i}{dt} \right) + k (x_{i+1}+x_{i-1}-2x_i) + f_i(t)\,,
\ee
where $f_i(t)$ is the external force applied during the driving phase.
In principle, a configuration of the spring-block model is defined by the list of positions $(x_0,x_1,\dots,x_N)$. However, two configurations 
$(x_0,x_1,\dots,x_N)$ and $(x_0+b,x_1+b,\dots,x_N+b)$ differing by a global translation $b$ can be considered as equivalent, because the dynamics given in
Eq.~(\ref{eq:dyn:spring-block}) is invariant under such a translation.
We thus rather characterize a configuration $\mC$ of the system by the list of spring elongations, $\mC=(\xi_1,\xi_2,\dots,\xi_N)$ with $\xi_i = x_i-x_{i-1}- l_0$ ($l_0$ being the rest length of the springs).

\subsection{Statistics of blocked configurations}

As mentioned above, a blocked configuration is defined by the set of conditions
\be \label{eq:cdt:blocked:force}
k |\xi_{i+1}-\xi_i|<\mu mg \quad (i=1,\dots,N-1),
\ee
corresponding to the fact that the sum of the forces exerted by springs is, in absolute value, less than the weight $mg$ times the friction coefficient $\mu$.
This condition can be reformulated by introducing the same length scale $a=\mu mg/k$ as in Eq.~(\ref{eq:def:a}) for the frictional harmonic oscillator,
leading to a condition equivalent to Eq.~(\ref{eq:cdt:blocked:force}),
\be \label{eq:cdt:blocked:a}
|\xi_{i+1}-\xi_i| < a \quad (i=1,\dots,N-1).
\ee
According to Edwards postulate, the distribution $P(\mC)$ of microscopic configurations is given by Eq.~(\ref{eq:Edwards:dist:E}),
where the energy $E(\mC)$ is the total elastic energy,
\be
E(\xi_1,\dots,\xi_N) =  \sum_{i=1}^N \frac{1}{2} k \xi_i^2\,,
\ee
and the indicator function $\mathcal{F}(\mC)$ can be written as
\be \label{eq:FC:SB}
\mathcal{F}(\xi_1,\dots,\xi_N) = \prod_{i=1}^{N-1} \Theta(a-|\xi_{i+1}-\xi_i|),
\ee
where $\Theta$ is again the Heaviside function equal to $\Theta(x)=1$ for $x\ge 0$ and to $\Theta(x)=0$ for $x<0$.
Eq.~(\ref{eq:FC:SB}) is a formal and compact way to express the list of conditions given in Eq.~(\ref{eq:cdt:blocked:a}).

The Edwards probability distribution $P(\mC)$ thus reads
\be
P(\xi_1,\ldots,\xi_N) = \frac{1}{Z} \exp\!\left[ -  \sum_{i=1}^N \frac{k \xi_i^2}{2T_{\rm eff}} \right]
\prod_{i=1}^N \Theta(a - |\xi_{i+1}-\xi_i|)
\label{eq:PC:spring-block}
\ee
with $Z$ a normalization factor defined as
\be \label{eq:def:Z:SB}
Z = \int d\xi_1\dots d\xi_N \,\exp\!\left[ -  \sum_{i=1}^N \frac{k \xi_i^2}{2T_{\rm eff}} \right]
\prod_{i=1}^N \Theta(a - |\xi_{i+1}-\xi_i|)
\ee
and playing the role of a partition function.
As before, the effective temperature $T_{\rm eff}$ is at this stage an auxiliary parameter that cannot be measured directly in a numerical simulation, and one needs to find a relation connecting $T_{\rm eff}$ to the energy density $\ve = \la E\ra/N$ which is a measurable observable.

Due to formal analogies between the Edwards distribution (\ref{eq:Edwards:dist:E}) and the usual equilibrium canonical distribution, similar relations hold between energy and temperature.
With the notation $\beta = T_{\rm eff}^{-1}$, the average energy is obtained similarly to the equilibrium case as
\be \label{eq:ElnZ:SB}
\langle E \rangle = -\frac{\partial \ln Z}{\partial \beta}\,,
\ee
as can be checked by a direct calculation.
The effective partition function $Z$ can be evaluated using a transfer operator technique. This method generalizes the transfer matrix technique classically used to solve for instance the one-dimensional Ising model \cite{Chaikin} by replacing the transfer matrix by an infinite dimensional operator that can be handled using numerical methods. The interested reader is referred to Ref.~\cite{Gradenigo15} for details.
An alternative method, that we now describe, consists in using an approximation that leads to analytically tractable calculations.
The basic idea is to replace the `door' function $\Theta(a-|\Delta\xi |)$ appearing in Eq.~(\ref{eq:FC:SB}) by a Gaussian function $\exp[-(\Delta\xi)^2/2a^2]$ with the same width $a$.

\begin{figure}[t!]
\centering\includegraphics[width=8cm]{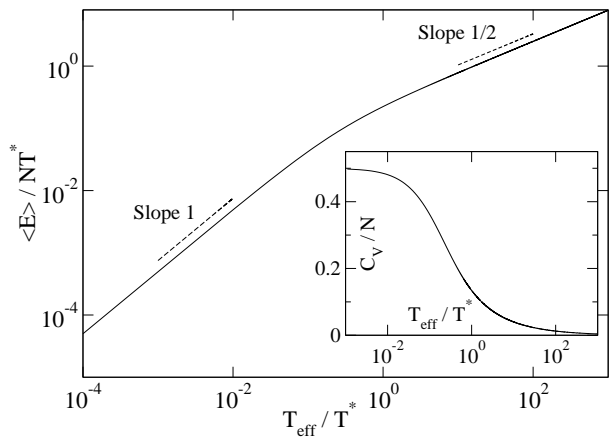}
\caption{Dimensionless energy density $\la E\ra/NT^*$ in the spring-block model, as a function of the reduced temperature $T_{\rm eff}/T^*$. The low- and high-temperature regimes are emphasized. Inset: specific heat $C_V/N$ versus $T_{\rm eff}/T^*$, showing a slow decrease from the classical value $1/2$ at low temperature to zero at high temperature.}
\label{fig-energy-SB}
\end{figure}

Under this approximation, the partition function $Z$ defined in Eq.~(\ref{eq:def:Z:SB}) can be rewritten as a Gaussian multidimensional integral
\be
Z = \int d\xi_1\dots d\xi_N \, e^{-\mathcal{H}(\xi_1,\dots,\xi_N)}
\ee 
where the quantity $\mathcal{H}(\xi_1,\dots,\xi_N)$ plays the role of an effective quadratic Hamiltonian,
\be
\mathcal{H}(\xi_1,\dots,\xi_N)
= \sum_{i=1}^{N} \left[\frac{1}{2}\beta k \xi_i^2 + \frac{1}{2a^2} (\xi_{i+1}-\xi_i)^2\right],
\label{eq:eff-act}
\ee
using periodic boundary conditions $\xi_{N+i}\equiv \xi_i$.
In more formal terms,
\be \label{eq:Gaussian:integ}
Z = \int d\xi_1\dots d\xi_N \, e^{-\frac{1}{2}{\bm \xi}\cdot \mathbf{A} {\bm \xi}}
\ee
having introduced the vector ${\bm \xi}=(\xi_1,\dots,\xi_N)$ and the
symmetric matrix $\mathbf{A}$ defined as
\be
(\mathbf{A}{\bm \xi})_j = \beta k \xi_j + \frac{1}{a^2}
(2\xi_j-\xi_{j+1}-\xi_{j-1})\,.
\ee
The general formula for multidimensional Gaussian integrals like the one of Eq.~(\ref{eq:Gaussian:integ}) reads \cite{Chaikin}
\be \label{eq:mult:Gauss}
Z = (2\pi)^{N/2}\, (\mathrm{det} \mathbf{A})^{-1/2} 
\ee
where $\mathrm{det} \mathbf{A}$ is the determinant of the matrix $\mathbf{A}$, that may be obtained as the product of all eigenvalues of the matrix $\mathbf{A}$. Eigenvectors here correspond to Fourier modes $\xi_j^{(q)}=e^{\mathrm{i} qj}$
($\mathrm{i}^2=-1$), and the corresponding eigenvalue reads
\be \label{eq:def:lambdaq}
\lambda_q = \beta k + \frac{2}{a^2} (1-\cos q)
\ee
where $q=2\pi n/N$ ($n=0,\dots,N-1$).
One then finds
\be
\ln \mathrm{det} \mathbf{A} = \sum_{n=0}^{N-1}
\ln \left[ \beta k + \frac{2}{a^2} \left(1-\cos \frac{2\pi n}{N}\right)\right],
\ee
eventually yielding in the large $N$ limit, using Eq.~(\ref{eq:mult:Gauss}) and the increment $\Delta q=2\pi/N$ to turn the sum over $q$ into an integral,
\be
\ln Z = \frac{N}{2}\ln (2\pi) - \frac{N}{4\pi}
\int_0^{2\pi} dq \, \ln \left[ \beta k + \frac{2}{a^2} (1-\cos q)\right].
\ee
The average energy $\la E\ra$ then follows by differentiating $\ln Z$ with respect to $\beta$, according to Eq.~(\ref{eq:ElnZ:SB}).
The resulting expression of $\la E\ra$ may be written in the form
\be \label{eq:avE:SB}
\la E \ra = \frac{1}{2}NT_{\rm eff}\, \fs \! \left(\frac{T_{\rm eff}}{T^*}\right)
\ee
with
\be \label{eq:def:Tst:SB}
T^*=k a^2
\ee
[note the slightly different definition of $T^*$ with respect to Eq.~(\ref{eq:def:a})], and a function $\fs(u)$ defined as
\be
\fs(u) = \frac{1}{\pi} \int_{0}^{\pi} \frac{dq}{1+2u(1-\cos q)}
= \frac{1}{\sqrt{1+4u}}\,,
\ee
where the second equality uses an integration formula from Ref.~\cite{Gradshteyn}. The average energy $\la E \ra$ is plotted versus temperature in Fig.~\ref{fig-energy-SB} in rescaled form.
We thus obtain the two asymptotic regimes:
\bea
\label{eq:equip1:SB}
\langle E \rangle &\approx& \frac{1}{2} \, N T_{\rm eff}  \qquad\qquad\;\, {\rm for} \quad T_{\rm eff} \ll T^* ,\\
\label{eq:equip2:SB}
\langle E \rangle &\approx& \frac{1}{4} \, N \sqrt{T^* \,T_{\rm eff}}\,
\qquad {\rm for} \quad T_{\rm eff} \gg T^* .
\eea
Hence, one finds a low energy regime where the energy density $\ve=E/N$ is proportional to the effective temperature $T_{\rm eff}$ similarly to the equipartition relation valid at equilibrium, and a high energy regime where $\ve$ is proportional to $\sqrt{T_{\rm eff}}$.
At odds with the frictional oscillator model, the energy density $\ve$ does not saturate to a maximal value when increasing the effective temperature $T_{\rm eff}$. Despite the lack of an upper bound on the energy, the specific heat $C_V = d\la E\ra/dT_{\rm eff}$ goes to zero at high effective temperature,
as $C_V \sim 1/\sqrt{T_{\rm eff}}$ (see inset of Fig.~\ref{fig-energy-SB}). This slow decay of $C_V$ is to be compared with the faster decay $C_V \sim 1/T_{\rm eff}^2$ for the frictional harmonic oscillator model of Sec.~\ref{sec:frict:HO}.
As a result, the equivalent of the Dulong and Petit law also breaks down for the spring-block model at high temperature.

\subsection{Correlation length and critical point}

Beside effective thermodynamic properties like the average energy, it is also of interest to determine the extent of spatial correlation in the system as a function of the effective temperature. In one-dimensional equilibrium systems with local interactions, like the one-dimensional Ising model for instance \cite{Chaikin}, the correlation length diverges when the temperature goes to zero, which corresponds to a zero-temperature critical point.
Given the formal similarities between the Edwards probability distribution (\ref{eq:Edwards:dist:E}) and the Boltzmann-Gibbs probability distribution,
one might expect the spring-block model to have a zero-temperature critical point. We will see below that, quite unexpectedy, the spring-block model has a critical point at infinite effective temperature, due to the constraints imposed by frictional properties.

Let us define the correlation function $C_r$ of the elongations of springs separated by a distance $r$,
\be \label{eq:def:Cr}
C_r = \frac{1}{N} \sum_{j=1}^N \la \xi_j \xi_{j+r} \ra\,,
\ee
using again periodic boundary conditions.
Note that the distance $r$ is measured in numbers of springs, rather than as a geometric length.
To proceed further, it is convenient to introduce the discrete Fourier transform
$\hat{\xi}_q$ as
\be \label{eq:def:xiq}
\hat{\xi}_q = \frac{1}{N} \sum_{j=0}^{N-1} e^{-\mathrm{i}qj}\, \xi_j \,,
\ee
with $q=2\pi n/N$ ($n=0,\dots,N-1$).
Similarly, one defines the discrete Fourier transform $\hat{C}_q$ of the correlation function $C_r$,
\be
\hat{C}_q = \frac{1}{N} \sum_{r=0}^{N-1} e^{-\mathrm{i}qr}\, C_r\,.
\ee
Using Eqs.~(\ref{eq:def:Cr}) and (\ref{eq:def:xiq}), one finds
$\hat{C}_q=\la | \xi_q |^2 \ra$.
Since Fourier modes diagonalize the matrix $\mathbf{A}$, standard properties of Gaussian multidimensional integrals \cite{Chaikin} lead to $\la | \xi_q |^2 \ra=2/\lambda_q$, where $\lambda_q$ is the eigenvalue defined in Eq.~(\ref{eq:def:lambdaq}).
One thus finds for small $|q|$ (assuming $N$ to be large):
\be \label{eq:Cq}
\hat{C}_q = 2 \left(\beta k+\frac{q^2}{2a^2}\right)^{-1}.
\ee
In the following, we approximate at large $N$ the discrete Fourier transform by a continuous Fourier transform.
Noticing that the continuous Fourier transform of an exponential correlation function
\be \label{eq:Cr}
C(r) = C_0 \, e^{-|r|/\ell}
\ee
reads
\be
\hat{C}(q) = \frac{2 C_0 \ell}{1+(q\ell)^2}
\ee
we identify from Eq.~(\ref{eq:Cq}) the expression
$\ell = \sqrt{T_{\rm eff}/(2T^*)}$ of the correlation length,
using the definition (\ref{eq:def:Tst:SB}) of $T^*$.

Note that the correlation length $\ell$ is dimensionless because it is measured in numbers of springs rather than as a geometric length.
To approximately convert it to a geometric length, one may simply multiply it by the rest length $l_0$ of the springs.

The continuous approximation of the Fourier transform used above is meaningful only when the correlation length is much larger than one, which implies $T_{\rm eff}\gg T^*$.
Given that in this regime the energy density also scales as $\ve \sim \sqrt{T_{\rm eff}}$ according to Eq.~(\ref{eq:equip2:SB}), we end up with the simple scaling relation $\ell \sim \ve$.
One thus concludes that the model exhibits a critical point at infinite effective temperature, or infinite energy density.
This unexpected property may be interpreted as follows. In the high energy regime, spring extensions $\xi_i$ are typically much larger than the length $a$
characterizing blocked configurations according to Eq.~(\ref{eq:cdt:blocked:a}).
Since the probability distribution is restricted to blocked configurations,
condition (\ref{eq:cdt:blocked:a}) is satisfied, with
both $|\xi_i|$ and $|\xi_{i+1}|$ typically much larger than $a$. This implies that $\xi_i$ and $\xi_{i+1}$ are often nearly equal, and are thus strongly correlated.
The decorrelation of the spring elongations $\xi_i$ and $\xi_{i+r}$ occurs only over distances $r\gtrsim \ell$.

\section{Suggested projects for students}
\label{sec:projects}

Several numerical projects of varying levels of difficulty are described below.

\subsection{Numerical simulations of the frictional harmonic oscillator}

The simplest numerical project consists in simulating the dynamics of a frictional harmonic oscillator given in Eq.~(\ref{eq:dyn:HOfrict}).
As the frictional harmonic oscillators described in Sec.~\ref{sec:frict:HO} are non-interacting, simulating a single oscillator is sufficient.
This can be done by discretizing with small time steps $\Delta t$ the differential equations obeyed by the position $x $ and velocity $v$ of the harmonic oscillator,
\be
\frac{d x}{dt} = v \,, \quad 
\frac{d v}{dt} = -k x - \mu g\, {\rm sign} (v) + f(t)\,,
\ee
where for the sake of simplicity we have set the mass $m$ to unity.
It is suggested to use an Euler integration scheme
\be
x(t+\Delta t) = x(t) + \frac{d x}{dt} \, \Delta t\,, \quad v(t+\Delta t) = v(t) + \frac{d v}{dt} \, \Delta t\,.
\ee
A driving protocol consists in repeatedly applying a force $f(t)$ for a duration $\tau$ and then let the system relax to a blocked configuration, before applying again a driving force.
A simple driving protocol may be to apply a force $f(t)$ that is constant during each driving period, with the same fixed amplitude $|f(t)|=f_0$ for all periods, but with a sign $f(t)=\pm f_0$ randomly drawn anew with equal probabilities at the beginning of each period. The amplitude $f_0$ has to be chosen significantly larger than the amplitude $\mu g$ of the friction force, say at least
$f_0 \gtrsim 5\mu g$. The overall intensity of the drive is essentially given by the product $f_0 \tau$ (this property has been checked in the spring-block model \cite{Gradenigo15}), and varying this driving intensity in different runs allows one to vary the average energy of the frictional harmonic oscillator.
The value of $\tau$ may be chosen of the order of $\tau\approx 100$.
One may also vary both $f_0$ and $\tau$ keeping the product $f_0 \tau$ constant to check that the average energy depends to a good approximation on the product 
$f_0 \tau$ only.

A subtle point, specific to systems subjected to dry friction, is that the particle starts to move only when the spring force overcomes the threshold value $\mu mg$. In the frictional oscillator model, this condition is equivalent to $|x_i| < a$ [see Eq.~(\ref{eq:def:a}) for the definition of $a$].
To initialize the dynamics, one may choose at random some initial conditions for the position $x$ and velocity $v$. If the condition $|x_i| \le a$ is satisfied, the particle does not move. If instead $|x_i| > a$, the particle starts to move according to the dynamical equation (\ref{eq:dyn:HOfrict}).
Motion goes on until the velocity $v$ becomes equal to zero, and the particle is at rest. Then the above condition to start motion applies again.

A natural goal of the simulation may be the measurement of the average energy density 
\be \label{eq:empirical:en:HO}
\ve=\frac{1}{KN} \sum_{j=1}^K \sum_{i=1}^N \frac{1}{2}k x_i^2(t_j)
\ee
where $t_j$, $j=1,\dots,K$ are a set of equidistant times satisfying $t_R = t_1 < t_2 < \dots < t_K = t_{\rm max}$, with $t_{\rm max}$ the maximal time of the simulation, and $t_R$ a relaxation time chosen such that memory of the initial condition is lost after an initial transient of duration $t_R$.

\subsection{Numerical simulations of the spring-block model}

A second project consists in integrating numerically the dynamics
Eq.~(\ref{eq:dyn:spring-block}) of the spring-block model,
which is both algorithmically and computationally more demanding.
The number $N$ of springs may be chosen according to the available computational facilities and to the amount of time dedicated to the project. The value of $N$ is a priori arbitrary, but typical value may range from a few units (e.g., $N=5$) to a few thousands.
The time-discretization of the dynamics given in Eq.~(\ref{eq:dyn:spring-block}) follows the same line as above, and the start and stop conditions need to be carefully implemented for each block.

A driving protocol consists in repeatedly applying a force $f_i(t)$ for a duration $\tau$ and then let the system relax to a blocked configuration, before applying again a driving force.
Two distinct driving protocols may be used. The first protocol is similar to the one used for the frictional harmonic oscillator: a force $f_i(t)$ with fixed magnitude $|f(t)|=f_0$ and a sign randomly chosen for each block at the beginning of each driving period is applied.
Alternatively, a second driving protocol consists in applying a force $f_i(t)=f_0$ on a fraction $\rho$ ($0<\rho<1$) of randomly chosen blocks. Namely,
for each block, a force $f_0$ is applied with probability $\rho$, otherwise no force is applied. This random choice is performed at the beginning of each driving period.
A simple choice is for instance $\rho=0.5$, but it may be of interest to test different values of $\rho$ in the numerical simulations to see how much the results depend on the value of $\rho$.

Like for the frictional harmonic oscillator, one may measure the average energy density 
\be
\ve=\frac{1}{KN} \sum_{j=1}^K \sum_{i=1}^N \frac{1}{2}k \xi_i^2(t_j)
\ee
with similar definitions of the time $t_j$ as in Eq.~(\ref{eq:empirical:en:HO}).
In addition, a more involved quantity to be measured is the correlation length of spring elongations, which requires to simulate larger system sizes, say at least $N=100$.
This correlation length is obtained by measuring the correlation function
\be
C(r) = \frac{1}{K N} \sum_{j=1}^K \sum_{i=1}^N \xi_i(t_j) \xi_{i+r}(t_j)\,,
\ee
for $r=0,\dots,N/2$ (assuming $N$ to be even).
Evaluating the correlation function $C(r)$ for different driving intensities, one can determine the corresponding correlation length $\ell$ from the relation
$C(\ell)=e^{-1}\, C(0)$. This procedure consistently allows for the correct determination of the correlation length for an exponential decay,
$C(r)=C(0)\, e^{-r/\ell}$, but does not assume the decay to be exponential. For consistency, it is useful to check that the different curves collapse when plotting $C(r)$ as a function of $r/\ell$ for different driving intensities \cite{Gradenigo15}.
It is suggested to use rather strong driving intensities, $f_0 \tau$
in the range $[10^3,10^4]$, to ensure that the correlation length
is significantly larger than one.

\smallskip

\section{Conclusion}
\label{sec:conclusion}

We have discussed simple examples of applications of the Edwards postulate for the statistical description of systems of particles experiencing dry friction.
We have seen in particular that the restriction of the probability distribution to blocked configurations has a strong impact on the phenomenology of frictional systems with respect to non-frictional ones, like the emergence of an infinite-temperature critical point in the spring-block model.
We have also seen in the two models presented that the effective specific heat goes to zero in the high temperature limit.
In practice, one of the main difficulties of the Edwards theory precisely lies in the evaluation of the function $\mathcal{F}(\mC)$ characterizing blocked configurations. The models considered in the present paper, being either without interactions or with a one-dimensional geometry, manage to keep this difficulty at a reasonable level. More realistic models of granular piles need to face this difficulty and to find appropriate approximations, see e.g.~\cite{BHDC15} for a review.


\begin{thebibliography}{99}



\bibitem{Edwards}
S.~F. Edwards, R.~B.~S. Oakeshott,
\emph{Theory of powders},
Physica A \textbf{157}, 1080 (1989).

\bibitem{ME89}
A. Mehta, S.~F. Edwards,
\emph{Statistical mechanics of powder mixtures},
Physica A \textbf{157}, 1091 (1989).

\bibitem{BKVS00}
A. Barrat, J. Kurchan, V. Loreto, M. Sellitto,
\emph{Edwards' measures for powders and glasses},
Phys. Rev. Lett. \textbf{85}, 5034 (2000).

\bibitem{BHDC15}
D.~P. Bi, S. Henkes, K.~E. Daniels, B. Chakraborty,
\emph{The statistical physics of athermal materials},
Annu. Rev. Condens. Matter Phys. \textbf{6}, 63 (2015).


\bibitem{BE03}
R. Blumenfeld and S.~F. Edwards,
\emph{Granular entropy: Explicit calculations for planar assemblies},
Phys. Rev. Lett. \textbf{90}, 114303 (2003).

\bibitem{BE06}
R. Blumenfeld and S.~F. Edwards,
\emph{Geometric partition functions of cellular systems: Explicit calculation of the entropy in two and three dimension},
Eur. Phys. J. E \textbf{19}, 23 (2005).

\bibitem{BSWM08}
C. Briscoe, C.~M. Song, P. Wang, and H.~A. Makse,
\emph{Entropy of Jammed matter},
Phys. Rev. Lett. \textbf{101}, 188001 (2008).

\bibitem{WSJM11}
P. Wang, C.~M. Song, Y.~L. Jin, and H.~A. Makse,
\emph{Jamming II: Edwards' statistical mechanics of random packings of hard spheres},
Physica A \textbf{390}, 427 (2011).

\bibitem{APF14}
D. Asenjo, F. Paillusson, and D. Frenkel,
\emph{Numerical calculation of granular entropy},
Phys. Rev. Lett. \textbf{112}, 098002 (2014).




\bibitem{HHC07}
S. Henkes, C.~S. O’Hern, and B. Chakraborty, 
\emph{Entropy and Temperature of a Static Granular Assembly: An Ab Initio Approach},
Phys. Rev. Lett. \textbf{99}, 038002 (2007).

\bibitem{HC09}
S. Henkes and B. Chakraborty, 
\emph{Statistical mechanics framework for static granular matter},
Phys. Rev. E \textbf{79}, 061301 (2009).

\bibitem{BE09}
R. Blumenfeld and S.~F. Edwards,
\emph{On Granular Stress Statistics: Compactivity, Angoricity, and Some Open Issues},
J. Phys. Chem. B \textbf{113}, 3981 (2009).

\bibitem{BJE12}
R. Blumenfeld, J.~F. Jordan, and S.~F. Edwards, 
\emph{Interdependence of the Volume and Stress Ensembles and Equipartition in Statistical Mechanics of Granular Systems},
Phys. Rev. Lett. \textbf{109}, 238001 (2012).

\bibitem{BZBC13}
D.~P. Bi, J. Zhang, R.~P. Behringer, and B. Chakraborty,
\emph{Fluctuations in shear-jammed states: A statistical ensemble approach},
Europhys. Lett. \textbf{102}, 34002 (2013).






\bibitem{NKBJN98}
E.~R. Nowak, J.~B. Knight, E. Ben-Naim, H.~M. Jaeger, and S.~R. Nagel, 
\emph{Density Fluctuations In Vibrated Granular Materials},
Phys. Rev. E \textbf{57}, 1971 (1998).

\bibitem{SGS05}
M. Schr\"oter, D.~I. Goldman, and H.~L. Swinney, 
\emph{Stationary state volume fluctuations in a granular medium},
Phys. Rev. E \textbf{71}, 030301(R).

\bibitem{LCDB06}
F. Lechenault, F. da Cruz, O. Dauchot, and E. Bertin,
\emph{Free volume distribution and compactivity measurement in a bidimensional granular packing},
J. Stat. Mech. P07009 (2006).

\bibitem{NRRCD09}
S. McNamara, P. Richard, S. de Richter, G. Le Ca\"er, and R. Delannay,
\emph{Measurement of granular entropy},
Phys. Rev. E \textbf{80}, 031301 (2009).



\bibitem{KM02}
J. Kurchan and H. Makse,
\emph{Testing the thermodynamic approach to granular matter with a numerical model of a decisive experiment},
Nature \textbf{415}, 614 (2002).

\bibitem{M04}
P.~T. Metzger,
\emph{Granular contact force density of states and entropy in a modified Edwards ensemble},
Phys. Rev. E \textbf{70}, 051303 (2004).

\bibitem{MD05}
P.~T. Metzger and C.~M. Donahue,
\emph{Elegance of disordered granular packings: A validation of Edwards' hypothesis},
Phys. Rev. Lett. \textbf{94}, 148001 (2005).

\bibitem{PCN06}
M. Pica Ciamarra, A. Coniglio, and M. Nicodemi,
\emph{Thermodynamics and Statistical Mechanics of Dense Granular Media},
Phys. Rev. Lett. \textbf{97}, 158001 (2006)

\bibitem{BK15}
V. Becker and K. Kassner,
\emph{Protocol-independent granular temperature supported by numerical simulations},
Phys. Rev. E \textbf{92}, 052201 (2015).


\bibitem{SL03}
Y. Srebro and D. Levine,
\emph{The Role of Friction in Compaction and Segregation of Granular Materials},
Phys. Rev. E \textbf{68}, 061301 (2003).


\bibitem{McQuarrie}
D.~A. McQuarrie, \textit{Statistical Mechanics} (Harper and Row, 1976).

\bibitem{Hill}
T.~L. Hill, \textit{An Introduction to Statistical Thermodynamics}
(Dover, 1986).

\bibitem{Chandler}
D. Chandler, \textit{Introduction to Modern Statistical Mechanics}
(Oxford, 1987).

\bibitem{Ashcroft}
N.~W. Ashcroft and N.~D. Mermin,
\emph{Solid State Physics}
(Saunders College Publishing, 1976).



\bibitem{BK67}
R. Burridge and L. Knopoff,
\emph{Model and theoretical seismicity},
Bull. Seismol. Soc. Am. {\bf 57}, 341 (1967). 

\bibitem{CL89}
J.~M. Carlson and J.~S. Langer, 
\emph{Mechanical model of an earthquake fault},
Phys. Rev. A {\bf 40}, 6470 (1989).

\bibitem{BPG11}
B. Blanc, L.-A. Pugnaloni, and J.-C. G\'eminard, 
\emph{Creep motion of a model frictional system},
Phys. Rev. E \textbf{84}, 061303 (2011).

\bibitem{Gradenigo15}
G. Gradenigo, E.~E. Ferrero, E. Bertin, J.-L. Barrat,
\emph{Edwards thermodynamics for a driven athermal system with dry friction},
Phys. Rev. Lett. \textbf{115}, 140601 (2015).



\bibitem{Chaikin}
P.~M. Chaikin and T.~C. Lubensky,
\emph{Principles of Condensed Matter Physics}
(Cambrigde University Press, Cambridge, 1995).

\bibitem{Gradshteyn}
I.~S. Gradshteyn and I.~M. Ryzhik,
\emph{Tables of Integrals, Series, and Products},
Fifth Edition (Academic Press, London, 1994).



\end{thebibliography}

\end{document}